\documentclass[12pt]{iopart}	
\pdfoutput=1 
\usepackage{color}
\definecolor{darkblue}{rgb}{0.,0.,0.4}
\definecolor{darkred}{rgb}{0.5,0.,0.}
\usepackage[USenglish]{babel} 
\usepackage[T1]{fontenc}
\usepackage[ansinew]{inputenc}
\usepackage{iopams}  
  \expandafter\let\csname equation*\endcsname\relax
  \expandafter\let\csname endequation*\endcsname\relax
\usepackage{amsmath}
\usepackage{lmodern} 
\usepackage[pdftex,colorlinks=true,linkcolor=darkblue, citecolor=darkred,urlcolor=blue]{hyperref}

\usepackage{graphicx}
\usepackage{epstopdf} 

\begin{document} 
\title[Heterodyne KPFM of gold and the Casimir force]{The effect of patch potentials in Casimir force measurements determined by heterodyne Kelvin probe force microscopy}

\author{Joseph L. Garrett$^{1,2}$, David Somers$^{1,2}$,
Jeremy N. Munday$^{2,3}$}

\address{$^1$ Department of Physics, College Park, MD 20742, USA}
\address{$^2$ Institute for Research in Electronics and Applied Physics, College Park, MD 20742}
\address{$^3$ Department of Electrical and Computer Engineering, College Park, MD 20742}

\begin{abstract}
	
Measurements of the Casimir force require the elimination of electrostatic interactions between the surfaces. However, due to electrostatic patch potentials, the voltage required to minimize the total force may not be sufficient to completely nullify the electrostatic interaction. Thus, these surface potential variations cause an additional force, which can obscure the Casimir force signal. In this paper, we inspect the spatially varying surface potential (SP) of e-beamed, sputtered, sputtered and annealed, and template stripped gold surfaces with Heterodyne Amplitude Modulated Kelvin Probe Force Microscopy (HAM-KPFM). It is demonstrated that HAM-KPFM improves the spatial resolution of surface potential measurements compared to Amplitude Modulated Kelvin Probe Force Microscopy (AM-KPFM).  We find that patch potentials vary depending on sample preparation, and that the calculated pressure can be similar to the pressure difference between Casimir force calculations employing the plasma and Drude models.
\end{abstract}



\submitto{\JPCM}

\maketitle

\section{Introduction}

	Patch potentials are predicted to cause a pressure between conductive surfaces and to introduce a small systematic error into Casimir force measurements \cite{Speake2003}. The Casimir force has been measured in various setups including torsion pendulums \cite{Lamoreaux1997}, atomic force microscopes \cite{Mohideen1998}, and microelectromechanical systems \cite{Chan2001,Decca2005}. Even in early measurements, patch potentials were suspected to cause an additional attractive force \cite{Lamoreaux1997}, though it was not directly observed. In more recent measurements, the force-minimizing voltage ($V_m$) was found to be distance dependent \cite{Kim2008} and was attributed to patch potentials.  
	
	Attempts have been made to quantify the pressure from patch potentials between parallel plates \cite{Speake2003} and its effect on Casimir force measurements \cite{Kim2010}. A perplexing situation has arisen in the field of Casimir physics where some experimental data has supported a theory based on use of the plasma model for conductivity \cite{Decca2005}, while other data has supported a Drude model \cite{Sushkov2011a}. The competition between the Drude and plasma models has focused attention on patch potentials \cite{Behunin2012,Behunin2012a}, as they may be a path to resolving the conflicting evidence from different experiments. 

	Electrostatic patch potentials complicate many other measurements as well \cite{Kim2010a}. Recent examples include increased noise in isoelectronic gravitation measurements \cite{behunin2014}, heating in ion traps \cite{Hite2013}, metal whiskers that short electonic circuitry \cite{karpov2014}, barrier height modification for Rydberg atom ionization \cite{Carter2011,Pu2010}, and limited conductivity in graphene on SiO$_2$ \cite{Burson2013}. In some cases, Kelvin probe force microscopy (KPFM) is used to measure the surface potential (SP) in order to diagnose experimental artifacts originating from electrostatic patches.
	Similarly, amplitude-modulated (AM) KPFM has been used to investigate patches originating from the crystal structure of copper \cite{Gaillard2006}. The potential contrast increased upon decreasing relative humidity, a phenomenon which suggested that in ambient conditions, the average patch size grows, but the voltage difference between patches decreases \cite{Behunin2012a}. 
	
	Recently, Kelvin probe techniques have been used to measure the SP of gold in Casimir force experiments \cite{Garcia-Sanchez2012a,Behunin}. The measurement of the SP on a gold-coated silicon nitride membrane \textit{in situ} with a modification of frequency modulated (FM) KPFM is reported in \cite{Garcia-Sanchez2012a}. The measured force agreed with that calculated in \cite{Kim2010} for two samples. The measurement, however, was limited in spatial resolution by size of the probe (4 mm), and was not able to resolve the smallest patches. Another recent experiment \cite{Behunin} measured the SP on a gold plate in a nitrogen environment, which had been used previously in vacuum-based Casimir force measurements \cite{Decca2007,Decca2007a}, and estimated the sphere/plate equivalent electrostatic pressure assuming both surfaces were covered with the measured potential. The estimated pressure was over an order of magnitude less than the discrepancy between the measured pressure and the pressure calculated by the Drude model. 
	
	In this paper, we first discuss our implementation of KPFM and show that it has sufficient resolution to observe patch potentials on e-beam deposited, sputtered, sputtered and annealed, and template-stripped gold (TSG) surfaces. Secondly, we analyze our SP measurements in light of a recent patch model \cite{Behunin2012}, and show that patch potentials would likely increase the measured force between these materials by an amount of the same order of magnitude as the plasma-Drude difference.

\section{Methods}

	\subsection{Kelvin Probe Force Microscopy}

\begin{figure}[h]
			\centering
			\includegraphics[width=.8\textwidth]{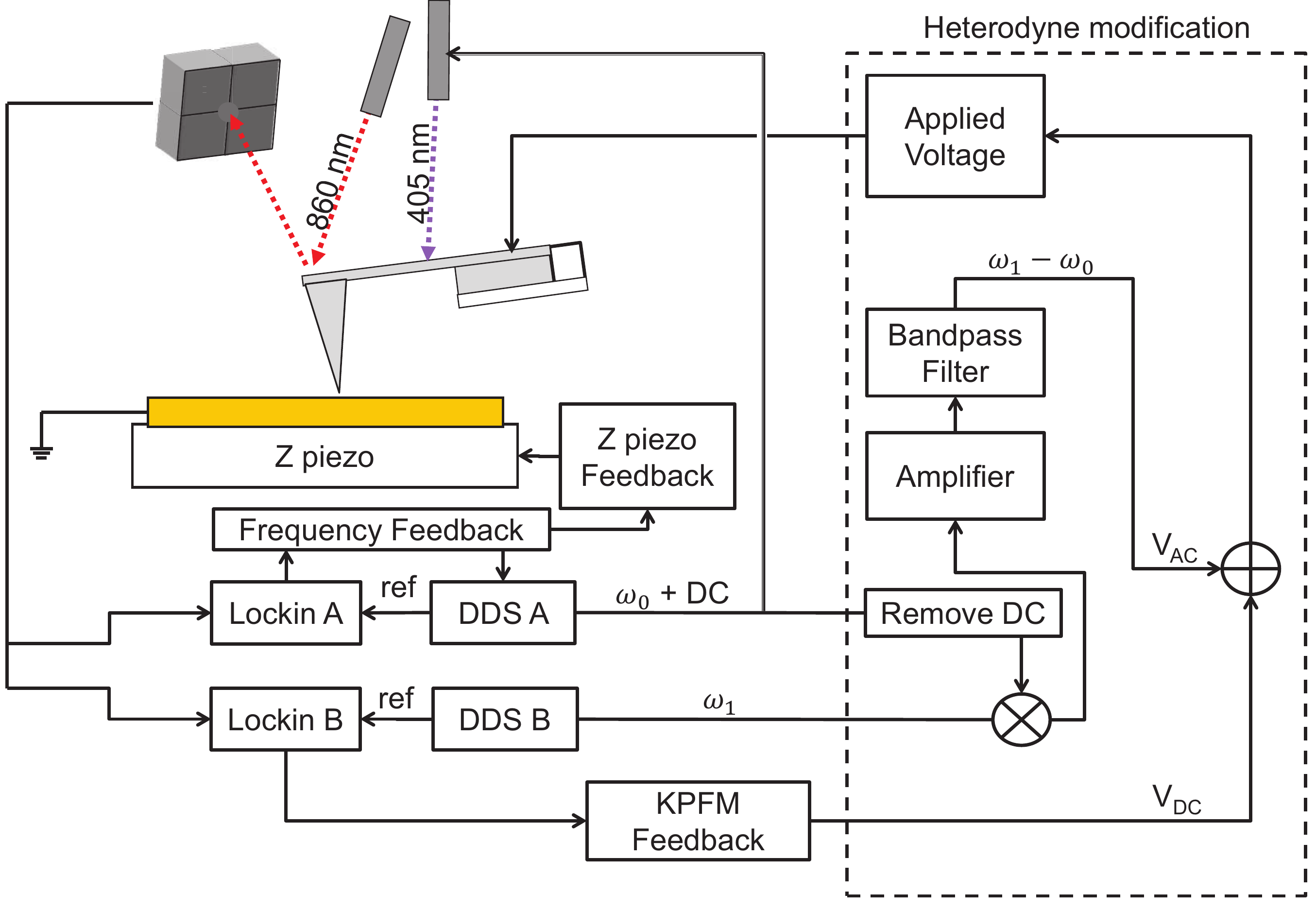}
			\caption{The left side of the figure shows the AFM set up for HAM-KPFM. The cantilever is thermally driven with a 405 nm laser, rather than a piezoelectric column (as in typical AM-KPFM). The right side of the figure shows the heterodyne modification. The signals from each lock-in are combined with a frequency multiplier before being amplified, filtered, and sent to the cantilever. }
			\label{fig:HAMKPFMfeedback}
		\end{figure}

	In AM-KPFM, an alternating voltage ($V_{AC}$) drives cantilever oscillations. A feedback loop minimizes the oscillations by applying a DC voltage ($V_{DC}$) to the probe to negate the contact potential difference (CPD) between the probe and the surface (defined here as $V_{0} = V_{tip}- V_{surface}$). This method has been used to measure SP for two decades \cite{Nonnenmacher1991}, but suffers from low spatial resolution relative to FM-KPFM \cite{Glatzel2003,Zerweck2005}. A recent variant, HAM-KPFM \cite{Sugawara2012} increases the spatial resolution by removing cantilever-induced capacitive artifacts \cite{Ma2013}, and is implemented on our AFM (Cypher, Asylum Research) with an external frequency multiplier (Minicircuits ZAD-8+) and a homemade bandpass filter/amplifier (see Figure ~\ref{fig:HAMKPFMfeedback}).  

	Platinum-coated cantilevers (NSC36/Pt-B, MikroMasch) with nominal spring constant of 2 N/m, a tip radius < 35 nm, $Q_{1}\approx 220$,  and a first eigenmode $\frac{\omega_{1}}{2 \pi}\approx$130 kHz are used. All of our measurements are in ambient conditions. Relative humidity is allowed to vary with the environment ($29\pm 5\%$) and the temperature is kept at $30.0\pm 0.1^{\circ}$C.
	
	Because HAM-KPFM is derived from AM-KFPM, we first discuss the original technique as an introduction and for comparison. The probe-surface interaction is modeled as a capacitor. The energy between a metal surface and a metal probe is $U = \frac{C V^2}{2}$,
	where C is the capacitance and V is the tip-surface potential difference. If $V$ is independent of tip-sample height, the force on the probe is:
	\begin{align}
	F_{z} &= -\frac{dC}{dz}\Big|_{z=d}\frac{V^2}{2},
	\end{align}	
	where $z$ is the vertical coordinate direction and $d$ is the instantaneous tip-sample separation. In AM-KPFM, an AC voltage $V_{AC}$ is applied to the cantilever so that:
	\begin{align}
	\label{eq:force}
	F_{z} &= F_{z, static} + F_{z, fun} + F_{z, harm},
	\end{align}
	where
	\begin{align}
	\label{eq:static}
	F_{z, static} &= -\frac{1}{2}\frac{dC}{dz}\Big|_{d}\Bigg[\frac{V_{AC}^2}{2}\ + (V_{0}+V_{DC})^2\Bigg],\\
	\label{eq:fun}
	F_{z, fun} &=-\frac{dC}{dz}\Big|_{d}V_{AC}(V_{0}+V_{DC})\text{cos}(\omega t),\\
	F_{z, harm} &= -\frac{1}{4}\frac{dC}{dz}\Big|_{d}V_{AC}^2\text{cos}(2\omega t).
	\end{align}
	The DC voltage $V_{DC}$ is applied to the probe in order to minimize the oscillation signal from $F_{z, fun}$. The minimizing voltage is $V_{DC} = - V_{0} = V_{surface}-V_{tip}$.
	
	For AM-KPFM, we use a two-pass technique \cite{Jacobs1997}. On the first pass, the cantilever measures the topography of the surface in non-contact amplitude-modulated mode. The cantilever is mechanically oscillated just above its resonance frequency, $\omega_{1}$, with $h\approx$ 12 nm, where $h$ is the time-averaged tip-sample separation. The vertical position of the sample is controlled with a voltage applied to a piezo beneath the sample. A feedback loop controls the height to maintain a constant amplitude of excitation, and consequently a constant $h$. 	
	On the second pass, the cantilever is raised 8 nm above the original scan height ($h\approx$ 20 nm), while a voltage oscillating at $\omega_{1}$ is applied to the probe and the sample is grounded. Any DC voltage difference between the probe and the surface excites the cantilever at the same frequency (see equation \ref{eq:fun}). A lock-in amplifier filters the signal and a feedback loop minimizes the in-phase component by applying a DC voltage to the probe. 
	
	In HAM-KPFM, topography and surface potential are imaged simultaneously. The position of the cantilever above the surface is controlled using a frequency modulated (FM) method \cite{Albrecht1991} with a negative frequency shift (for scan size 1 $\mu\text{m}^{2}$, $\frac{\Delta \omega}{2 \pi}=$--50 Hz) and an amplitude of 20 $\pm$ 2 nm at $\omega_{1}$. Increasing the magnitude of our negative frequency shift improves the voltage resolution but hinders the stability of the measurement. We use the largest magnitude frequency shift that allows stability to be maintained for a particular scan speed, as determined experimentally. 
	
The voltage resolution is improved in HAM-KPFM by increasing the proportion of the signal due to the tip relative to the beam of the cantilever. The signal is amplified by the second resonance of the cantilever. Simultaneous with topographic imaging, an AC voltage is applied at a frequency $\omega_{bp} = \omega_{2} - \omega_{1}$, where $\omega_{2}$ is the second eigenmode frequency of the cantilever \cite{Sugawara2012} ($Q_{2}\approx 450$, $\frac{\omega_{2}}{2 \pi} \approx$ 950 kHz), and $\omega_{bp}$ is the frequency of the bandpass filter. The instantaneous separation between the tip and the sample is:
		\begin{align}
			\label{eq:oscillation}
			d = A\text{cos}(\omega_{1} t) + h,
		\end{align}
	where $A$ is the amplitude of oscillation at $\omega_{1}$, and $h$ is the period-averaged height from the surface to the tip, as before. The capacitance gradient is:
		\begin{align}
		\label{eq:capacitance expand}
		\frac{dC}{dz}\Big|_{d} \approx& \frac{dC}{dz}\Big|_{h}  +A\frac{d^{2}C}{dz^{2}}\Big|_{h}\text{cos}(\omega_{1} t),\\
		\notag
		\end{align}
		in the limit $A\ll h$. The electrostatic force due to excitation at the fundamental frequency (equation \ref{eq:fun}) becomes:
		\begin{align}
		\label{eq:Hamexpand}
		F_{z,fun} \approx& -\Bigg[\frac{dC}{dz}\Big|_{h}	+
		A \frac{d^2C}{dz^2}\Big|_{h} \text{cos}(\omega_{1} t)\Bigg]V_{AC}(V_{0}+V_{DC})\text{cos}(\omega_{bp} t).
		\end{align}
		The oscillatory force at the 2nd eigenmode, $\omega_{2}$, is:
		\begin{align}
		\label{eq:Hamexpand}
		F_{z,\omega_{2}} \approx&
		-\frac{A}{2} \frac{d^2C}{dz^2}\Big|_{h} V_{AC}(V_{0}+V_{DC}) \text{cos}(\omega_{2} t).
		\end{align}
	Again, a feedback loop minimizes the oscillation at $\omega_{2}$ by applying a DC voltage to the probe (note: the applied DC voltage is the CPD). The excitation measured in the heterodyne method depends on the second derivative of the tip-sample capacitance, which has a stronger distance dependence than the first derivative \cite{Hudlet1998,Glatzel2003}. This dependence decreases the contribution of the cantilever to the potential measurement, improving its spatial resolution \cite{Glatzel2003,Sugawara2012}. For example, \cite{Zerweck2005} reports a factor of 8 improvement in the SP resolution of a KCl-Au boundary when using FM-KPFM (which also depends on $\frac{d^2C}{dz^2}$) compared to AM-KPFM (which depends on $\frac{dC}{dz}$). Similar improvement is expected for HAM-KPFM.
				
	Unless indicated otherwise, our scans are square, with 1024 x 1024 pixels. Lines are raster scanned at 0.3 Hz, with $V_{AC}$ = 1 V. The scans are all referred to by the length of a side here onwards. The size of our scans is limited by the stability of our FM-topography feedback loop. Because we are interested in variations and not the mean of the potential, we remove the mean from each scan, which also removes the effect of $V_{tip}$.

	\subsection{Surface Preparation}
	
	We prepared e-beam deposited, sputtered, sputtered and annealed, and template stripped gold (TSG) samples in order to determine how patch potentials vary with preparation. Gold was deposited to 100 nm thickness for all samples. One sample was e-beam deposited (Denton) at 3 $\mu $Torr with a 10 nm Cr sticking layer onto a polished silicon wafer, epoxied to a metallic puck pior to deposition (EPO-TEK E4110). The other samples were sputtered (AJA International, Inc.) in a 2.5 mTorr argon environment onto a polished silicon wafer. Two were sputtered with a 20 nm Cr sticking layer, and their silicon substrates were epoxied to the pucks. The third was sputtered onto a silicon wafer directly, and a puck was epoxied to the exposed Au surface so that it could be template stripped \cite{Blackstock2003,Ederth2000}. 
	All three sputtered samples were heated for 6 hours at 80 $^{\circ}$C to solidify the epoxy. One of the samples with a Cr sticking layer was annealed at 200 $^{\circ}$C in ambient atmosphere for an additional 3 hours. The TSG sample was mechanically removed from the silicon immediately before scanning.
	
	\subsection{Pressure due to Patch Potentials}
	\subsubsection{Theory}
	
	A method has been developed to describe the pressure due to patch potentials in terms of the patch correlation function between parallel plates \cite{Speake2003}, which is given by:
	\begin{align}
		\label{eq:correlationfunction}
		C_{i,j}(\vec{x}) &= \iint d^2\vec{x'} \: V_{i}(\vec{x'})\: V_{j}(\vec{x'} + \vec{x}),
	\end{align}
	where $\vec{x}$ and $\vec{x}'$ are the spatial coordinates, and $V_{i,j}$ are the potentials, average removed, on each plate. The correlation function is converted to a radial form, $C(r)$, and is averaged over all angles at each position $r$. To calculate the force, the correlation function is transformed into k-space \cite{Behunin2012}:	
	\begin{align}
		\label{eq:radialcorr}
		C_{i,j}[k] = 2\pi\int^{\infty}_{0} dr \: r \: C_{i,j}(r) \: J_{0} (k r),
	\end{align}		
	where $J_{0}$ is a Bessel function.
	The pressure between two parallel plates resulting from patches is: 
	\begin{align}
		\label{eq:radialcorr}
		P^{Patch}_{1,2}(d) = &\frac{\epsilon_{0}}{4\pi}\int^{\infty}_{0} dk \frac{k^3}{\text{sinh}(kd)^2} \big[C_{1,1}[k] + C_{2,2}[k] -2 C_{1,2}[k]\text{cosh}(kd)\big],
	\end{align}	
	where $d$ is the distance between the two plates \cite{Speake2003,Behunin2012}. Here, we consider the pressure between plates with the same autocorrelation, but vanishing cross-correlation, so that:
	
	\begin{align}
	\label{eq:patchpressure}
	P^{Patch}(d)\approx &\frac{\epsilon_{0}}{2\pi}\int^{\infty}_{0} dk \frac{k^3}{\text{sinh}(kd)^2}C[k].
	\end{align}
	
	\subsubsection{Calculation}
	
	Once the potential is determined by KPFM, the mean of the potential image is subtracted. The x-y autocorrelation function of the SP is calculated, binned into a radial autocorrelation function, and normalized to the number of pixels summed over at each distance. Because our scan size is limited, the $C(r)$ is truncated for $r>\frac{L}{2}$ (by analogy to the Nyquist-Shannon sampling theorem \cite{Sonka2007}). This procedure may slightly reduce the calculated pressure, but at longer distances $C(r)$ is sampled less and so is not representative of the sample. The radial autocorrelation function is numerically integrated to calculate $C[k]$. The pressure between parallel plates is calculated from equation \ref{eq:patchpressure}. 
		\begin{figure}[h]
			\centering
			\includegraphics[width=.7\textwidth]{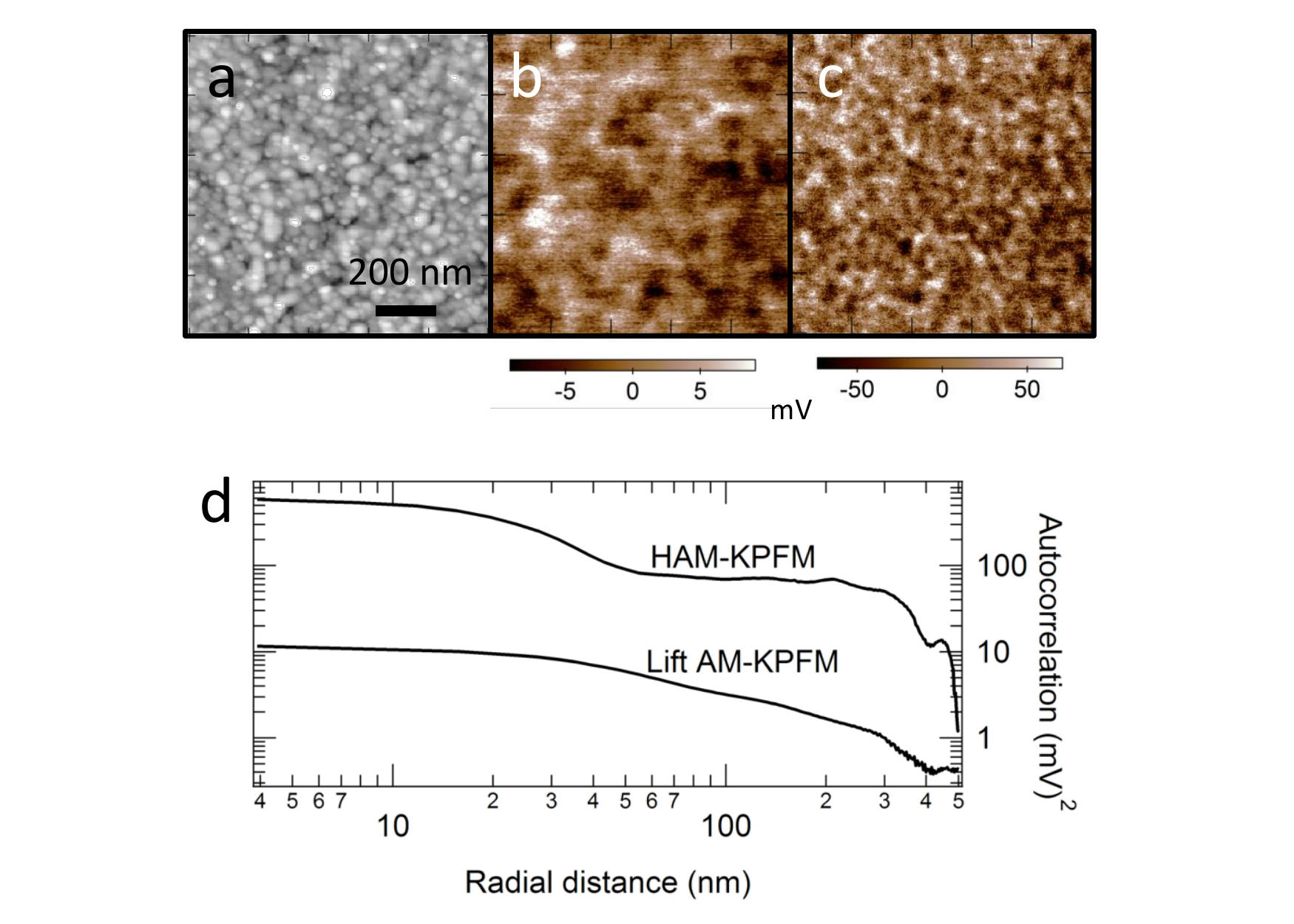}
			\caption{Comparison between HAM-KPFM and AM-KPFM. The scale of the sputtered Au topography is $\pm$ 3 nm, as acquired in AM mode (a). The SP acquired with AM-KPFM lift mode at 20 nm above the surface (b) shows some potential variations, but they are broad, and only differ by about 20 mV.  The SP acquired with HAM-KPFM reveals a higher spatial resolution (c), and, consequently, more potential contrast. The autocorrelation functions illustrate the difference between the two methods (d). 
			}
			\label{fig:CompAMLHAMscans}
		\end{figure}
\section{Results}

	\subsection{Calibrating HAM-KPFM}
	
	We measure the potential and topography of a 1x1 $\mu$m area, much larger than the patches observed, on the sputtered Au sample in order to compare the spatial resolution of AM- and HAM-KPFM (Figure \ref{fig:CompAMLHAMscans}). The scans, 256 x 256 pixels each, are acquired at 0.5 Hz with $V_{AC}$ = 1 V. The drift between the measurements is calculated from the topography of each and found to be about 6 nm. The potential acquired in HAM-KPFM (\ref{fig:CompAMLHAMscans}b) shows more overall contrast than the AM-KPFM lift mode image ($\approx180$ mV vs. $\approx20$ mV, \ref{fig:CompAMLHAMscans}c) and has increased spatial resolution, although some of the same patches can be seen in both. HAM-KPFM resolves patches on the surface more clearly and thus allows for a better estimate of the patch potential pressure. 

\begin{figure}[h]
			\centering
			\includegraphics[width=.99\textwidth]{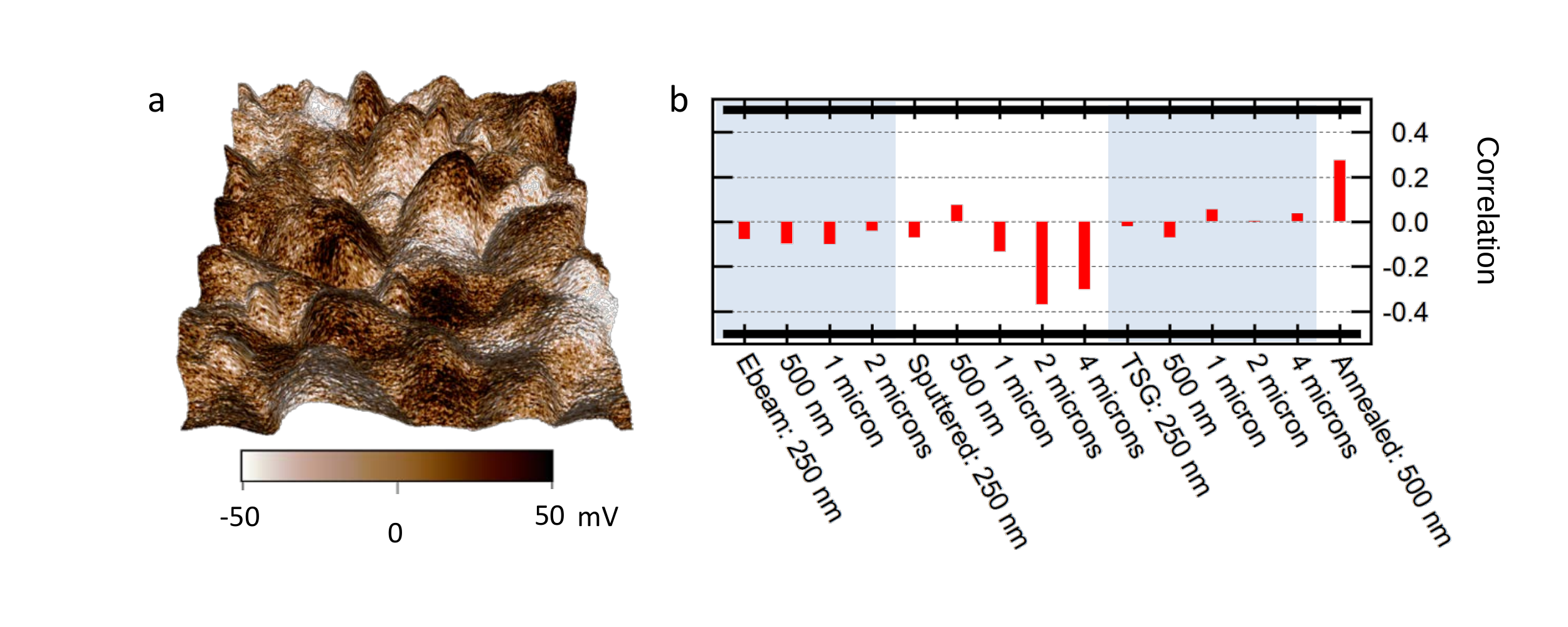}
			\caption{The AFM scan of a sputtered gold surface (250 nm) has topographical peaks at both high and low potentials (a) (total topographic range is 5.7 nm over the scanned area). The cross correlation between surface potential and topography falls below the limit used in \cite{Behunin} (thick lines at top and bottom) and takes on both positive and negative values depending on the sample and scan (b). 
			}
			\label{fig:InsignificantTopographicalImprint}
		\end{figure}

	Because it is known that the measured potential can be coupled to topography or vary as a function of distance in Kelvin probe measurements \cite{Zerweck2005,Liscio2008}, we investigated the effect of topography on HAM-KPFM. Figure \ref{fig:InsignificantTopographicalImprint}a shows a 250 nm scan of a sputtered Au surface with potential (mean removed) overlaid on topography. The image shows peaks with both positive and negative potential, and valleys with both positive and negative potential. This data suggests that topography does not leave a noticeable imprint on HAM-KPFM measurements. 
	To further investigate the effect of topography on the potential, we calculated their cross-correlation (Figure \ref{fig:InsignificantTopographicalImprint}b). While noting that a correlation between potential and topography does not imply a topographical artifact in the potential, nor does its absence imply the absence of a topographical artifact, it is a useful way to compare our measurements to \cite{Behunin}, which designated 0.5 as a maximum acceptable topographical-potential correlation. We found that the topography-potential correlation in our measurements varied significantly from scan to scan, but remained below 0.5 for all. Furthermore, the correlations took on both positive and negative values depending on the scan size and sample, which would be unlikely to occur if there were a systematic topographical imprint.
	
	\begin{figure}[h]
		\centering
		\includegraphics[width=1\textwidth]{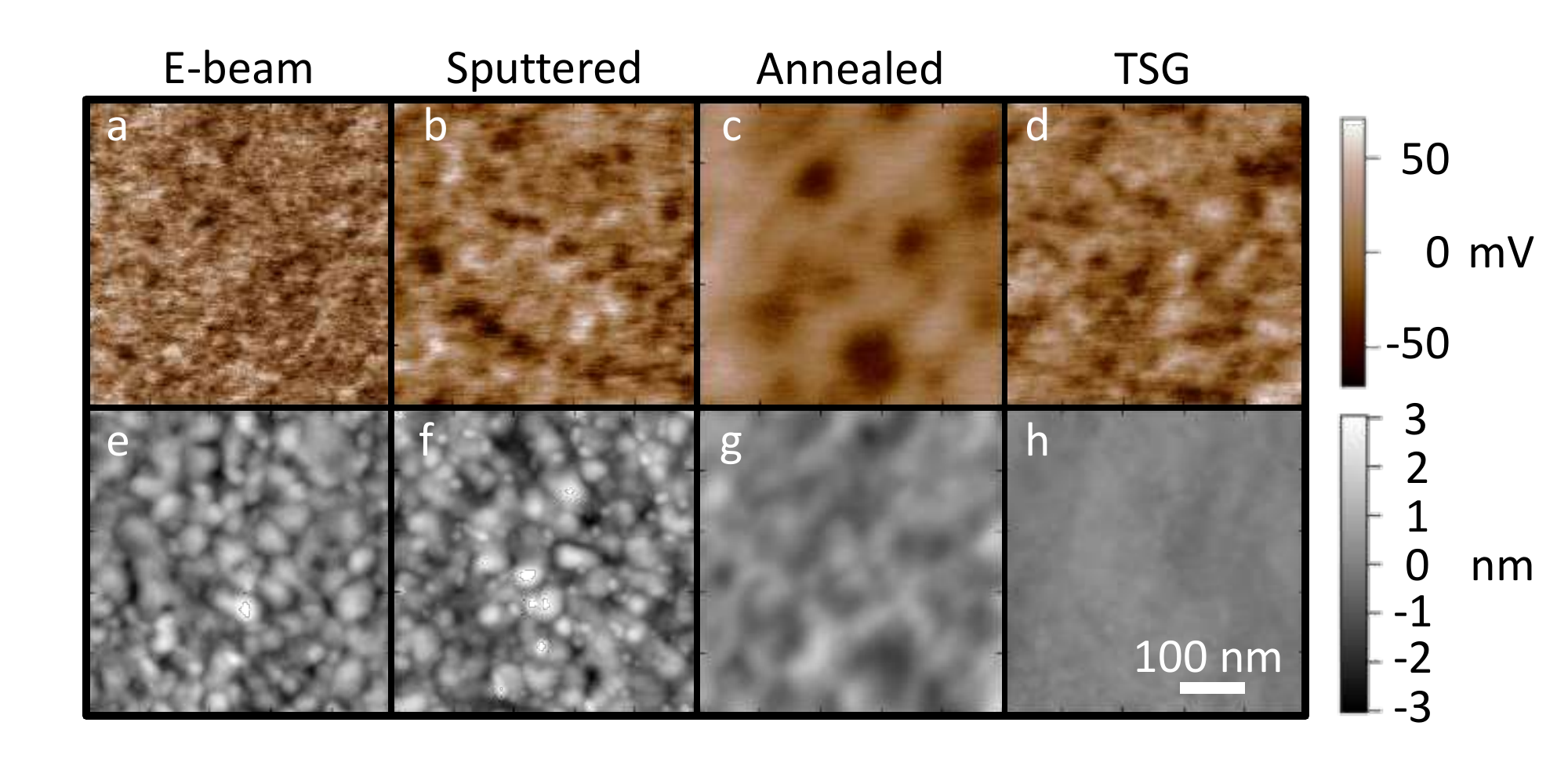}
		\caption{Surface potential of gold for e-beamed, sputtered, annealed, and template stripped gold (a-d) respectively. The topography recorded simultaneously (e-h). (color online)
		}
		\label{fig:500nm4surfaces}
	\end{figure}
	
	\subsection{Surface Potential}
	
	The spatial distribution of the surface potential we observe differs significantly from sample to sample; however, the amount of the variation remains similar. Figure \ref{fig:500nm4surfaces} shows 500 nm scans of the surfaces, both the potential and topography. The e-beamed sample has the smallest patches, while the patches on the sputtered and TSG samples are of similar size. The largest patches are found on the annealed sample.
	
	The topography also varies significantly. Both the e-beamed and sputtered samples have similar roughness, while the sputtered and annealed sample shows less topographical variation. The TSG sample is the smoothest, as expected \cite{Ederth2000}. 
	
\begin{figure}[h]
		\centering
		\includegraphics[width=.9\textwidth]{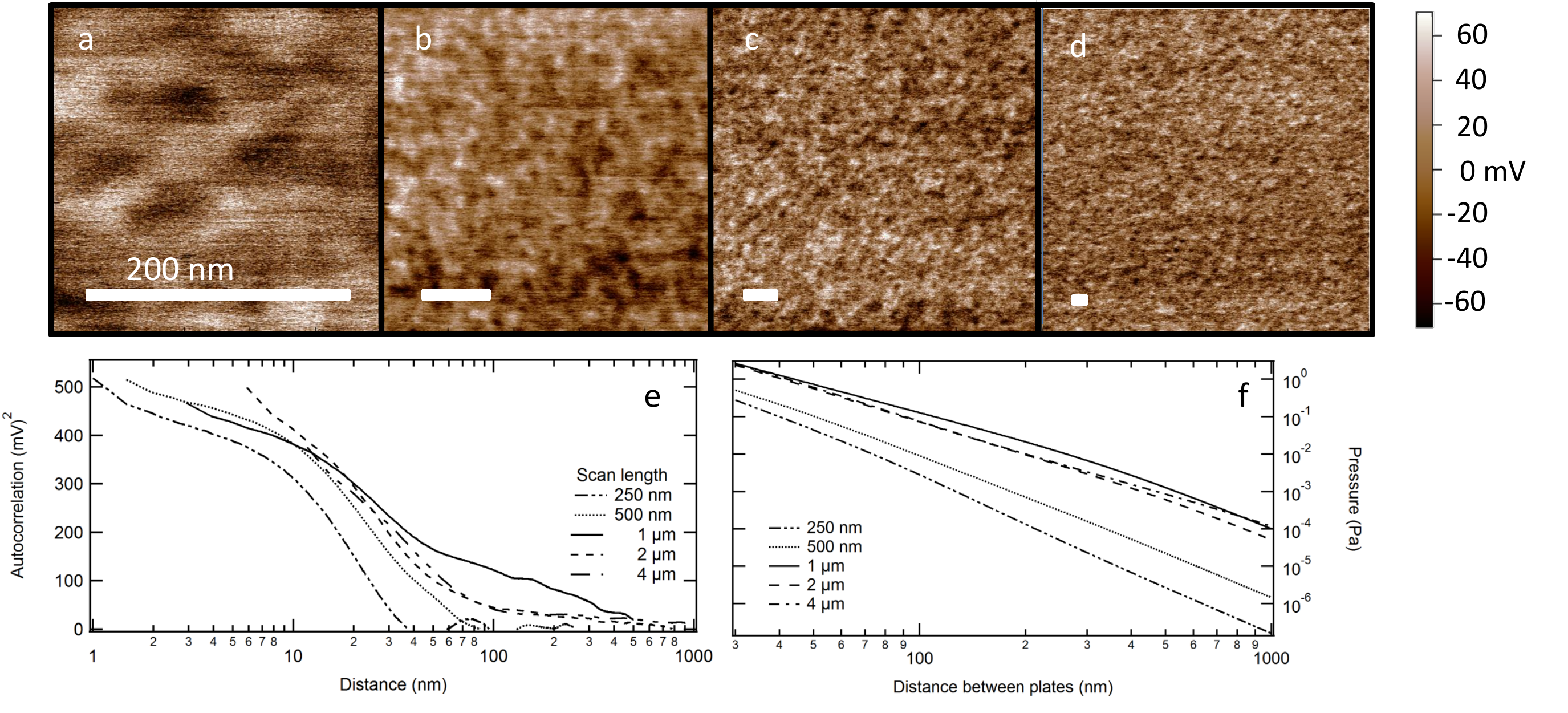}
		\caption{The measured patch potentials on a sputtered gold surface maintain the same general shape and distribution as the size of the scan is increased from 250 nm (a) to 1 $\mu$m (b), 2 $\mu$m (c), and 4 $\mu$m (d). However, the calculated autocorrelation function falls off much more quickly for increasing $r$ in small scans (e) which leads to a smaller calculated pressure (f).
		}
		\label{fig:SPsizescaling}
	\end{figure}

	To understand how scan size affects the calculated pressure between plates, we investigate how $C(r)$ changes with scan size. Here we focus on the sputtered sample. Figures \ref{fig:SPsizescaling} (a-d) and Figure \ref{fig:500nm4surfaces}b show scans of the sputtered gold sample ranging from 250 nm to 4 $\mu$m. All scans are 1024 x 1024 and recorded at 0.3 Hz. The frequency offset was chosen from - 40 Hz (4 $\mu$m) to - 70 Hz (250 nm), in order to maximize resolution while maintaining stability. The general patch shape and size are consistent as the scan size is varied. 
	
	In some of the scans there is an abrupt difference between scan lines, a phenomenon which has been attributed to charge transfer at the tip \cite{Weaver1991}. Because we are concerned with the potential of the surface and not of the tip, the scan is 0th order flattened (the average of each scan line is subtracted) to mitigate the effect of any change in $V_{tip}$. This procedure slightly decreases the amplitude of $C(r)$ for all the scans. However, for the smallest scans, it also introduces anticorrelations which were not seen in the larger scans. For this reason, we do not include any 500 or 250 nm scans in the pressure calculations of the subsequent sections. To observe the effect of 0th order flattening, compare the images in Figure \ref{fig:500nm4surfaces} to those in Figures \ref{fig:CompAMLHAMscans} and \ref{fig:SPsizescaling} (unflattened). A recent analysis of the spheres used in Casimir force measurements showed that flattening can remove low-frequency spectral data from an AFM image and affect the calculated autocorrelation function \cite{Sedmik2013}, particularly at long distances. Here, we remove $V_{tip}$ variations in order to avoid an overestimate of the pressure. 
	
\section{Discussion}

\begin{figure}[h]
	\centering
	\includegraphics[width=1\textwidth]{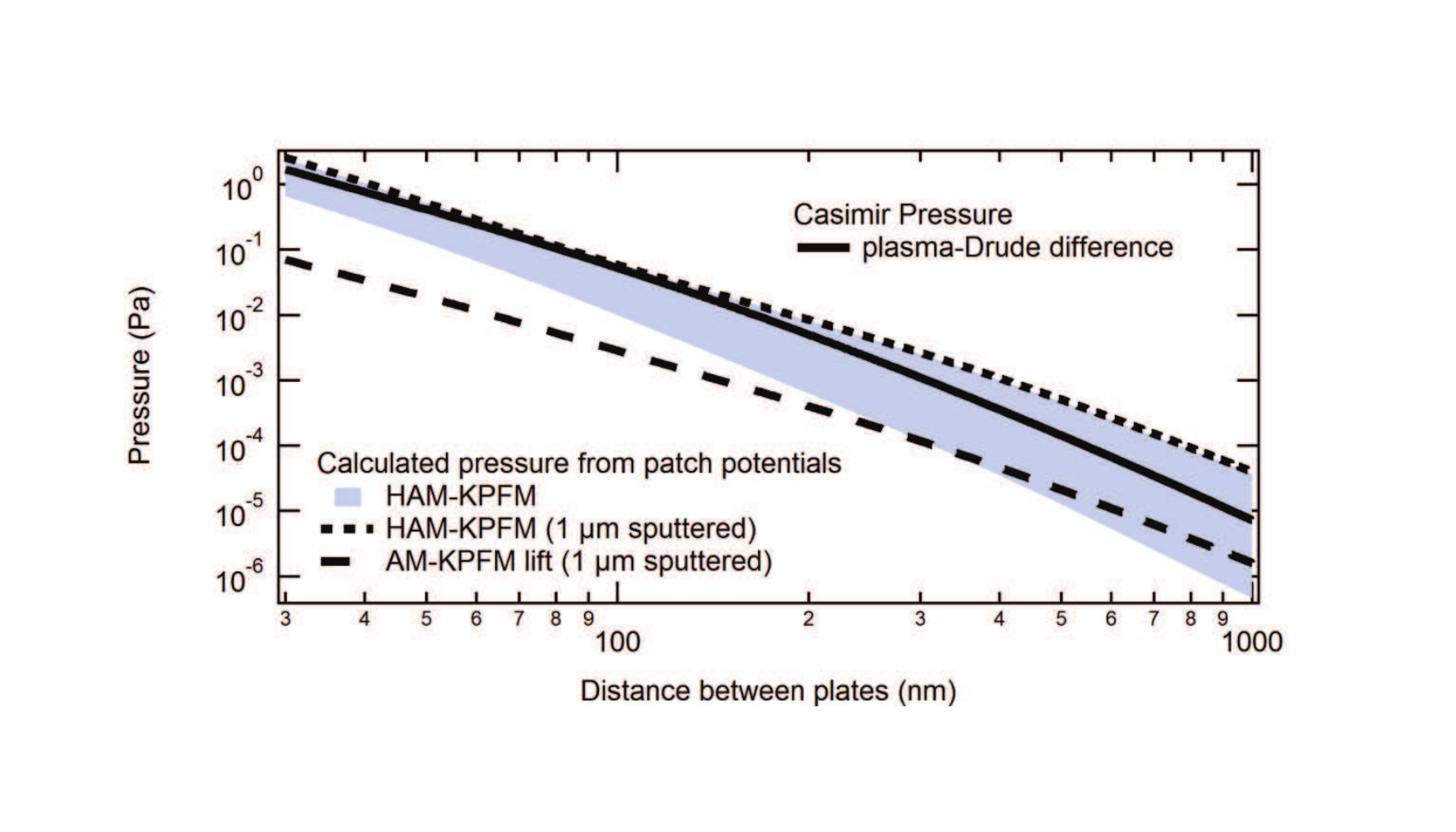}
	\caption{The pressure difference between the plasma and Drude models for the Casimir force ($P_{difference}=P_{plasma}-P_{Drude}$, solid black line) falls within the range of patch potential pressures calculated from the autocorrelation functions directly (blue band, all scans 1 $\mu$m$^2$ or greater). AM-KPFM (dashed line) predicts a much smaller force than HAM-KPFM (dotted line) for the same area (1 $\mu$m$^2$). All pressures displayed here are attractive.
	}
	\label{fig:Forces}
\end{figure}

\subsection{Comparing surface potential pressure to plasma and Drude models for the Casimir force}	

	We calculate the pressures from the 1, 2, and 4 $\mu$m scans and compare them to the plasma-Drude Casimir pressure difference. Before computing pressures, the surface potential measurements are 0th order flattened, as discussed above, and 3x3 median filtered in order to remove stochastic noise \cite{Sonka2007}. 

	The Casimir pressure for parallel gold plates is calculated at 300 K, using the plasma and Drude models with $\tilde{\gamma} = 0.035$ eV and $\tilde{\omega}_{p} = 9.0$ eV \cite{Lambrecht2000,Parsegian2006} without the addition of optical data at other frequencies, both for simplicity and because optical properties vary based on preparation \cite{Svetovoy2008}. Although most Casimir force experiments are performed in a sphere/plate geometry, the calculation here remains in a parallel plate formulation, as the proximity force approximation is not sufficient to convert the patch potential pressure to a sphere/plate geometry \cite{Behunin2012a}.

	The patch potential pressures computed from different samples varies by over an order of magnitude at a separation of 1 $\mu$m and by a factor of 4 at 30 nm (Figure \ref{fig:Forces}). The plasma-Drude pressure difference remains within the range of patch potential pressures at all distances computed here.

	\subsection{Modeling patch potentials for Casimir force measurements}
	The potential variation and patch size distribution can be obtained from our measurements and used in simple patch potential models when the exact SP is unknown. Here we compare our measurements to the quasi-local model \cite{Behunin2012} (QLM), which takes the patch size distribution and $V_{RMS}$ as inputs. In \cite{Behunin2012} it was suggested that in vacuum conditions, the grain size determines the patch potential size, while in ambient conditions, patch potentials result from adsorbates, and thus are larger in size but of less magnitude. The size of patches that we found in an image do not seem to correlate directly to the grain size of the material, as established from an AFM image. For example, in Figure \ref{fig:500nm4surfaces}, the e-beam and sputtered samples have very similar topography, while the length scales of their patch potentials differ.
		
		\begin{figure}[h]
			\centering
			\includegraphics[width=.8\textwidth]{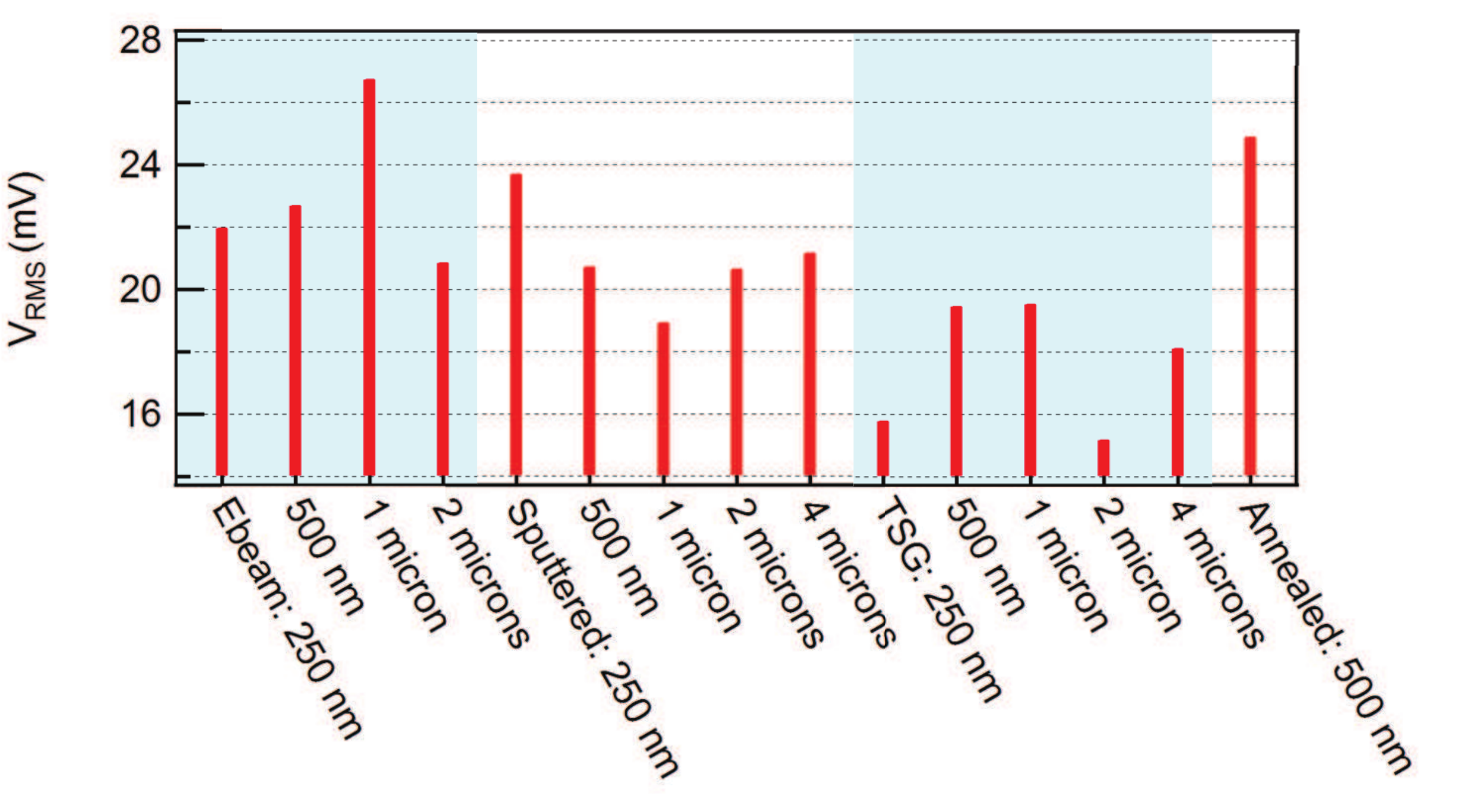}
			\caption{$V_{RMS}$ shows sample dependence, but generally falls between 15 and 28 mV, suggesting a range of $V_{RMS}$ to use when estimating the patch potential force from these samples.
			}
			\label{fig:Crs}
		\end{figure}
		
	 The surface potentials measured here, in ambient conditions, show variation on both small (30 - 100 nm) and large scales (300 - 1000 nm), although the latter variation is of less magnitude and more sensitive to flattening. The smallest patches are of the same size as the tip radius, so it is likely that still smaller patches exist on the surface. Both preparation and scan size influence $V_{RMS}$, but in general it falls between 15 and 28 mV (after a 3x3 median filter to remove noise) (Figure \ref{fig:Crs}). 
	
\begin{figure}[h]
				\centering
				\includegraphics[width=.8\textwidth]{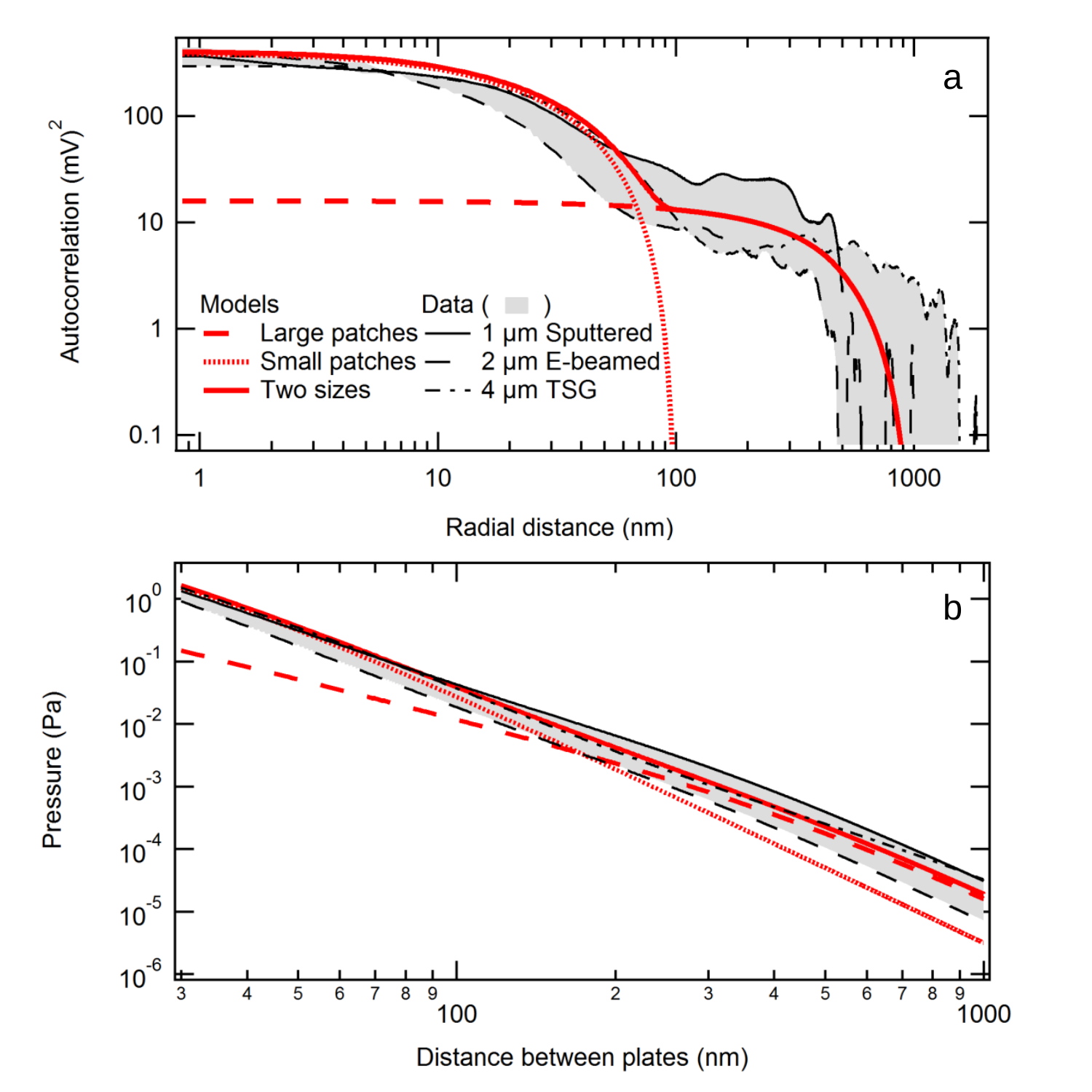}
				\caption{In the quasi-local model, neither small patches nor large patches alone match the $C(r)$ data from experiments; however, their combination, by equation \ref{eq:patches}, does (a). The parameters are, for small patches: $l_{min} = 10$ nm, $l_{max} = 100$ nm, $V_{RMS} = 20$ mV, while for large patches: $l_{min} = 500$ nm, $l_{max} = 1 \: \mu$m, $V_{RMS} = 4$ mV using the uniform size distribution as in \cite{Behunin2012}. Consequently, two patch sizes also better approximate the calculated pressure from the measured data as well (b). 
				(color online)}
				\label{fig:Crmodel}
			\end{figure}

	Two ranges of patch size are incorporated into the QLM to replicate the two observed scales of correlation. In \cite{Behunin2012}, to use the QLM, the patch size distribution is considered to be constant for $l_{min}<l<l_{max}$, where $l$ is the diameter of a patch. The calculated $C(r)$ is determined by the parameters $V_{RMS}$, $l_{min}$ and $l_{max}$. Here, multiple patch sizes are incorporated into the QLM by giving each different patch size range a different $V_{RMS}$. If the large and small patches are uncorrelated, 	
	\begin{equation}
		\label{eq:patches}
		C(r) = C^{large}(r) + C^{small}(r).
	\end{equation}	
		This model allows the creation of autocorrelation functions where long-range correlations are present (to allow for adsorbates and contamination often unavoidable), while still considering the short-range correlations which cause most of $V_{RMS}$ and gives a better approximation of the measured autocorrelation functions (Figure \ref{fig:Crmodel}a). 
		
	A simulation of patch potentials was used to conclude that patch potentials did not contribute to a measured force in \cite{Chang2012}. Because $V_{m}$ varied with distance in the simulation when patch potentials were present, the lack of a distance dependent $V_{m}$ led the authors to conclude the patch potential force was zero. The range of SP values ($\pm\:90$ mV) was similar to our observations, however, their spatial form differed. The potentials were modeled as square patches of side length $s$ on a grid with spacing $l>s$. Outside the squares, the SP took on a uniform value. In our experiment, patches varied continuously, and there were small long-distance correlations not present in the simulated model. Distance dependence of $V_{m}$ is suppressed when the area of interaction ($2\pi R d$, where $R$ is the sphere radius, and $d$ is the sphere/plate separation) is much larger than the average patch \cite{Kim2010,Lamoreaux2011,Behunin2012a}, here found to be about than 100 nm across. Thus, $V_{m}$ distance dependence may not be observed even if patches still contribute to the force for $d \gg \frac{r^2_{patch}}{2R}$, where $r_{patch}$ is the typical patch radius. 	

\section{Conclusions}

	Patch potentials became an important consideration in Casimir force measurements for two reasons: a distance-dependent $V_{m}$ \cite{Kim2008}, and the possibility that they might offer a route to resolve the plasma-Drude controversy \cite{Behunin2012}. Here we show that the plasma-Drude pressure difference falls within the range of pressures generated by patch potentials on several gold surfaces, as measured by HAM-KPFM, which differs from an earlier result using AM-KPFM \cite{Behunin}. Furthermore, we demonstrate an improved spatial resolution when using HAM-KPFM instead of AM-KPFM on the same area. 
	 
	We have shown that the size of patch potentials depends on sample preparation, but $V_{RMS}$ varies by less than a factor of 2 for all preparation techniques used. The quasilocal correlation model introduced by Behunin \cite{Behunin2012}, describes patches at one size scale well and can be adapted to a two-scale model, as we have done. A single KPFM scan may not be representative of a surface, particularly if the patch size is large. To know how patch potentials contribute to an experiment, it is necessary to know the exact, not statistical, potential at the location of closest measurement. Because the patch potential pressure is of the same order of magnitude as the pressure difference between the plasma and Drude models, HAM- or FM-KPFM should be performed on both the sphere and plate in the same environment as the Casimir force measurement, because environmental effects on patch potentials are still an open problem.
		 
We acknowledge the support of the Maryland NanoCenter and its FabLab and, in particular thank Jon Hummel and Tom Loughran for assistance depositing gold and acknowledge the University of Maryland for laboratory startup funds. We also would like to thank the scientists at Asylum Research for conversations about implementing HAM-KPFM.\\

\bibliographystyle{iopart_num}
\bibliography{PatchPotentials_Arxiv}

\end{document}